\documentclass[a4paper,11pt]{article}
\pdfoutput=1 

\usepackage{jheppub} 
\title{Electric and magnetic energy at axion haloscopes}

\author[a]{B. R. Ko,}
\author[a]{H. Themann,}
\author[a]{W. Jang,}
\author[a]{J. Choi,}
\author[a]{D. Kim,}
\author[a]{M. J. Lee,}
\author[b,1]{J. Lee,\note{Corresponding author.}}
\author[c]{E. Won,} 
\author[a,b]{and Y. K. Semertzidis\,}
\affiliation[a]{Center for Axion and Precision Physics Research,
  Institute for Basic Science (IBS), Daejeon 34141, Republic of Korea}
\affiliation[b]{Department of Physics, Korea Advanced Institute of
  Science and Technology (KAIST), Daejeon 34141, Republic of Korea}
\affiliation[c]{Department of Physics, Korea University, Seoul 02841,
  Republic of Korea}

\emailAdd{jhinhwan@kaist.ac.kr}

\abstract{
  We review a recent letter published in
  Phys. Rev. Lett. \textbf{116}, 161804 (2016) of which the main
  argument is that the mode dependent magnetic form factors at axion
  haloscopes depend on the position of the cavity inside the solenoid
  while the corresponding electric form factors do not. We, however,
  find no such dependence, which is also equivalent to the statement
  that the electric and corresponding magnetic energy stored in the
  cavity modes are the same regardless of the position of the cavity
  inside the solenoid. Furthermore, we extend the statement to the
  cases satisfying $\vec{\nabla}\times\vec{B}_{\rm external}=0$, where
  $\vec{B}_{\rm external}$ is a static magnetic field provided by a
  magnet at an axion haloscope. Two typical magnets, solenoidal and
  toroidal, satisfy $\vec{\nabla}\times\vec{B}_{\rm external}=0$, thus
  the electric and corresponding magnetic energy stored in the cavity
  modes are always the same in both cases. The energy, however, is
  independent of the position of the cavity in axion haloscopes with a
  solenoid, but depends on those with a toroidal magnet.
}


\begin{document}

\maketitle
\flushbottom

\section{Introduction}
The axion is a hypothetical particle that was proposed as a solution
to the strong $CP$
problem~\cite{strongCP1,strongCP2,strongCP3,strongCP4,strongCP5,strongCP6,strongCP7}. This
particle is massive, abundant, and very weakly interacting, making it
a promising candidate for cold dark matter if the axion mass is above
1 $\mu$eV/$c^2$~\cite{CDM1,CDM2,CDM3} and below 3
meV/$c^2$~\cite{SN1987_1,SN1987_2,SN1987_3,SN1987_4,SN1987_5}.

The axion search method proposed by Sikivie~\cite{sikivie}, also known
as the axion haloscope search, involves a microwave resonant cavity
with a strong static magnetic field that induces axion conversions
into microwave photons, where the electromagnetic energy from the
interaction inside the cavity is
\begin{equation}
  U_{a,em}=g^2_{a\gamma\gamma}\langle a^2\rangle c^2\epsilon_0 B_{\rm avg}^2 V C_{\rm mode},
  \label{EQ:AGG_ENERGY}
\end{equation}
where $g_{a\gamma\gamma}$ is the axion-photon coupling strength,
$\langle a^2\rangle$ is the mean-square of the spatially homogeneous
axion field, $a=a_0 e^{-i\omega_a t}$, with angular
frequency $\omega_a$, corresponding to the axion mass, and amplitude
$a_0$~\cite{HALO_AXION}, $c$ the speed of light
(thus, $c^2\epsilon_0=1/\mu_0$). $B^2_{\rm avg}$ is defined with
$B^2_{\rm avg} V\equiv\int\vec{B}^2_{\rm external}dV$, where
$\vec{B}_{\rm external}$ is a static magnetic field provided by
magnets at axion haloscopes, and $V$ is the volume of the cavity. The
form factor of the cavity depending on a resonant mode $C_{\rm mode}$
is
\begin{equation}
  C_{\rm mode}=\frac{\biggl(\int_{V} dV\sum\limits^3_{i=1} E_{i, {\rm mode}}B_i\biggr)^2}
    {B^2_{\rm avg} V\int_{V}dV\sum\limits^3_{i=1}\sum\limits^3_{j=1}
      E_{i, {\rm mode}} \epsilon_{ij} E_{j, {\rm mode}}},
  \label{EQ:FF.gen}
\end{equation}
where $B_i$ is the component of $\vec{B}_{\rm external}$, $E_{i, {\rm mode}}$
is the electric field component of a resonant mode, and
$\epsilon_{ij}$ is component of the relative dielectric tensor.
In the axion haloscopes to
date~\cite{haloscope1,haloscope2,haloscope3,haloscope4,haloscope5},
eq.~(\ref{EQ:FF.gen}) has been used to calculate the form factor of a
cylindrical cavity that is centered in and occupies the complete
volume of a solenoidal field.

Recently, a report~\cite{EMFF_HOJOO} pointed out that
eq.~(\ref{EQ:FF.gen}) actually corresponds only to electric energy
from axion to photon conversions inside the cavity $U_{a,e}$, thus
they referred to eq.~(\ref{EQ:FF.gen}) as the electric form factor
$C_E$. The report also pointed out that an implicit assumption that
$U_{a,e}$ and $U_{a,m}$ are the same has been made, where $U_{a,m}$ is
the magnetic energy from the conversion inside the cavity. Hence the
assumption results in $U_{a,em}=2U_{a,e}$, which, according to this
report, has been valid to date for a cylinder in a solenoid as
mentioned in the previous paragraph. Furthermore,
the report introduced the magnetic form factor of the cavity modes
$C_B$ or, equivalently, the magnetic energy stored in the cavity modes
of axion haloscopes and claimed that $C_B$ depends on the offset of a
cylindrical cavity from the center of the solenoid while the
corresponding electric form factors do not show such dependence. Their
claim is effectively equivalent to the statement that the electric
energy stored in the cavity modes is generally different from the
corresponding magnetic energy in axion haloscopes.

In this paper we show that there is no such dependence with
cylindrical axion haloscopes with solenoidal $B$ fields, which was
already recognized in Ref.~\cite{SANGJUN}. Furthermore, we also find
that as long as $\vec{\nabla}\times\vec{B}_{\rm external}=0$ is valid,
the electric and magnetic energy stored in the cavity modes are the
same for both a solenoid and a toroidal magnet.

\section{Maxwell's equations with axion modifications}
Our review starts from Maxwell's equations with modification due to
the axion field~\cite{MAXWELL_AXION}
\begin{eqnarray}
  \label{EQ:MMAXWELL}    
  \vec{\nabla}\cdot(\vec{E}-cg_{a\gamma\gamma}a\vec{B})
  &=&\frac{\rho_e}{\epsilon_0}, \\
  \nonumber
  \vec{\nabla}\cdot(\vec{B}-\frac{1}{c}g_{a\gamma\gamma}a\vec{E})
  &=&\mu_0 \rho_m, \\
  \nonumber
  \vec{\nabla}\times(\vec{E}-cg_{a\gamma\gamma}a\vec{B})
  &=&-\frac{\partial}{\partial t}(\vec{B}-\frac{1}{c}g_{a\gamma\gamma}a\vec{E})-\mu_0\vec{J}_{m}, \\
  \nonumber
  \vec{\nabla}\times(\vec{B}+\frac{1}{c}g_{a\gamma\gamma}a\vec{E})
  &=&\frac{1}{c^2}\frac{\partial}{\partial t}(\vec{E}-cg_{a\gamma\gamma}a\vec{B})+\mu_0\vec{J}_{e},
\end{eqnarray}
where $\rho_e$ and $\vec{J}_e$ are electric charge density and
current, $\rho_m$ and $\vec{J}_m$ are magnetic charge density and
current if magnetic monopoles exist\footnote{No experimental evidence
  for magnetic monopole yet, thus, $\rho_m=0$ and $\vec{J}_{m}=0$ in
  eq.~(\ref{EQ:MMAXWELL}) would be desirable.}. In
eq.~(\ref{EQ:MMAXWELL}), $\vec{E}$ and $\vec{B}$ are the electric and
magnetic fields that are not induced from the interaction ruled by
$g_{a\gamma\gamma}$, but participate in that
interaction~\cite{COMPLEX}. We will refer to the electric and magnetic
fields induced from the coupling as $\vec{E}_a$ and $\vec{B}_a$ which
are related to the second terms in parentheses in
eq.~(\ref{EQ:MMAXWELL}). In principle, $\vec{E}_a$ and $\vec{B}_a$ can
also interact with the axion field to induce additional axion to
photon conversions, but the interactions are proportional to
$g^2_{a\gamma\gamma}a^2$ of which the magnitude is
$\mathcal{O}(10^{-43})$, thus can be ignored in
eq.~(\ref{EQ:MMAXWELL}).

\section{Maxwell's equations with axions only}
We make the following simplifying assumptions
\begin{itemize}
\item vacuum, thus $\rho_e=\vec{J}_e=\rho_m=\vec{J}_m=0$, 
\item no electromagnetic radiation, thus $\vec{E}=0$, but $\vec{B}=\vec{B}_{\rm external}$, 
\end{itemize}
where $\vec{B}_{\rm external}=B_0\hat{z}$ or $\vec{B}_{\rm
  external}=\frac{B_0}{\rho}\hat{\phi}$, and $B_0$ is a constant. Note
that $z$, $\rho$, and $\phi$ refer to cylindrical coordinates. With
our simplifying assumptions and ignoring terms with
$g^2_{a\gamma\gamma}a^2\sim\mathcal{O}(10^{-43})$, both sides of the
first three equations in eq.~(\ref{EQ:MMAXWELL}) are zero. The
left-hand side of the last equation in eq.~(\ref{EQ:MMAXWELL}) becomes
$\vec{\nabla}\times\vec{B}_{\rm external}$ which goes to zero with
ideal solenoids and toroidal magnets, but the right-hand side becomes
\begin{equation}
-\frac{1}{c}g_{a\gamma\gamma}\vec{B}_{\rm external}\frac{\partial a}{\partial t},
\label{EQ:CURRENT}
\end{equation}
which is not zero in the presence of a time-varying axion
field. Equation~(\ref{EQ:CURRENT}) can be regarded as a time-varying
displacement current, $\mu_0\vec{J}_a$, induced from the variation of
axion field or a time-varying electric field induced from the
variation of the axion field, thus also can be equated to
$\frac{1}{c^2}\frac{\partial\vec{E}_a}{\partial t}$. Then, according
to Ampere-Maxwell law, eq.~(\ref{EQ:CURRENT}) should induce magnetic
field $\vec{B}_a$, which has to go to left-hand side of the last
equation in eq.~(\ref{EQ:MMAXWELL}). Having said that, the last
equation in eq.~(\ref{EQ:MMAXWELL}) with our simplifying assumptions
can be written as
\begin{eqnarray}
  \vec{\nabla}\times\vec{B}_a
  &=&-\frac{1}{c}g_{a\gamma\gamma}\vec{B}_{\rm external}\frac{\partial a}{\partial t} \\
  \nonumber
  &=&\mu_0\vec{J}_a\\
  \nonumber
  &=&\frac{1}{c^2}\frac{\partial\vec{E}_a}{\partial t}.
  \label{EQ:AMAXWELL}        
\end{eqnarray}
In the presence of $\vec{E}_a$ and $\vec{B}_a$, Maxwell's equations
satisfying our simplifying assumptions become
\begin{eqnarray}
  \label{EQ:AMAXWELL2}    
  \vec{\nabla}\cdot\vec{E}_a
  &=&0, \\
  \nonumber
  \vec{\nabla}\cdot\vec{B}_a
  &=&0, \\
  \nonumber
  \vec{\nabla}\times\vec{E}_a
  &=&-\frac{\partial\vec{B}_a}{\partial t}, \\
  \nonumber
  \vec{\nabla}\times\vec{B}_a
  &=&\frac{1}{c^2}\frac{\partial\vec{E}_a}{\partial t}.
\end{eqnarray}
Note that eq.~(\ref{EQ:AMAXWELL2}) is valid if
$\vec{\nabla}\times\vec{B}_{\rm external}=0$, which is the case
with axion haloscopes in solenoidal or toroidal magnetic fields. Note
also that eq.~(\ref{EQ:AMAXWELL2}) is the same form as the Maxwell's
equations with electromagnetic radiation only, meaning one can treat
$\vec{E}_a$ and $\vec{B}_a$ in the same way that has always been done
when treating the electric and magnetic fields associated with
radiation.

At axion haloscopes, we are actually interested in the energy stored
in a cavity. The source of this energy is from the time-varying
electric ($\vec{E}_{ac}$) and magnetic ($\vec{B}_{ac}$) fields induced
by the axion field. Note that $\vec{E}_{ac}$ and $\vec{B}_{ac}$ will
have to satisfy the boundary conditions for a given cavity at the
end. Then, $\vec{E}_{ac}$ and $\vec{B}_{ac}$ must satisfy Maxwell's
equations in eq.~(\ref{EQ:ACMAXWELL})
\begin{eqnarray}
  \label{EQ:ACMAXWELL}    
  \vec{\nabla}\cdot\vec{E}_{ac}
  &=&0, \\
  \nonumber
  \vec{\nabla}\cdot\vec{B}_{ac}
  &=&0, \\
  \nonumber
  \vec{\nabla}\times\vec{E}_{ac}
  &=&-\frac{\partial\vec{B}_{ac}}{\partial t}, \\
  \nonumber
  \vec{\nabla}\times\vec{B}_{ac}
  &=&\frac{1}{c^2}\frac{\partial\vec{E}_{ac}}{\partial t},
\end{eqnarray}
under our simplifying assumptions. Since $\vec{E}_{ac}$ and
$\vec{B}_{ac}$ oscillate with the frequency $\omega_a$,
eq.~(\ref{EQ:ACMAXWELL}) can be written as
\begin{eqnarray}
  \label{EQ:AMAXWELL3}    
  \vec{\nabla}\cdot\vec{E}_{ac}
&=&0, \\
  \nonumber
  \vec{\nabla}\cdot\vec{B}_{ac}
  &=&0, \\
  \nonumber
  \vec{\nabla}\times\vec{E}_{ac}
  &=&i\omega_a\vec{B}_{ac}, \\
  \nonumber
  \vec{\nabla}\times\vec{B}_{ac}
  &=&-\frac{i\omega_a}{c^2}\vec{E}_{ac}.
\end{eqnarray}

\section{Electric and magnetic energy at axion haloscopes}
One can get the relation between electric and magnetic energy from
eq.~(\ref{EQ:AMAXWELL3})
\begin{equation}
  \vec{E}_{ac}\cdot\vec{E}^*_{ac}=c^2\vec{B}_{ac}\cdot\vec{B}^*_{ac}-\frac{ic^2}{\omega_a}\vec{\nabla}\cdot(\vec{E}^*_{ac}\times\vec{B}_{ac}).
  \label{EQ:EMDENSITY_CAVITY}
\end{equation}
The second term in right-hand side of eq.~(\ref{EQ:EMDENSITY_CAVITY}) goes
to zero with the boundary conditions at the cavity surface because we
do not expect electromagnetic energy flow through the cavity surface
in an ideal case. Therefore, eq.~(\ref{EQ:EMDENSITY_CAVITY}) states
that the electric energy stored in a cavity mode
$U_{ac,e}\propto\vec{E}_{ac}\cdot\vec{E}^*_{ac}$ and the corresponding
magnetic energy stored in a cavity mode
$U_{ac,m}\propto\vec{B}_{ac}\cdot\vec{B}^*_{ac}$ are the same,
regardless of the position of the cavity inside the magnet as long as
$\vec{\nabla}\times\vec{B}_{\rm external}=0$.
In the absence of dielectric material in the system $U_{a,e}$ and
$U_{a,m}$ can be equated to $U_{ac,e}$ and $U_{ac,m}$,
respectively. Since we show $U_{ac,e}=U_{ac,m}$ regardless of the position
of the cavity inside the magnet as long as
$\vec{\nabla}\times\vec{B}_{\rm external}=0$, $U_{a,e}=U_{a,m}$ is
true, thus $C_E=C_B$ is also true regardless of the position of the
cavity inside the solenoid, which is different than the argument in
Ref.~\cite{EMFF_HOJOO}.
\begin{figure}[htbp]
  \centering
  \includegraphics[height=0.6\textwidth,width=0.9\textwidth]{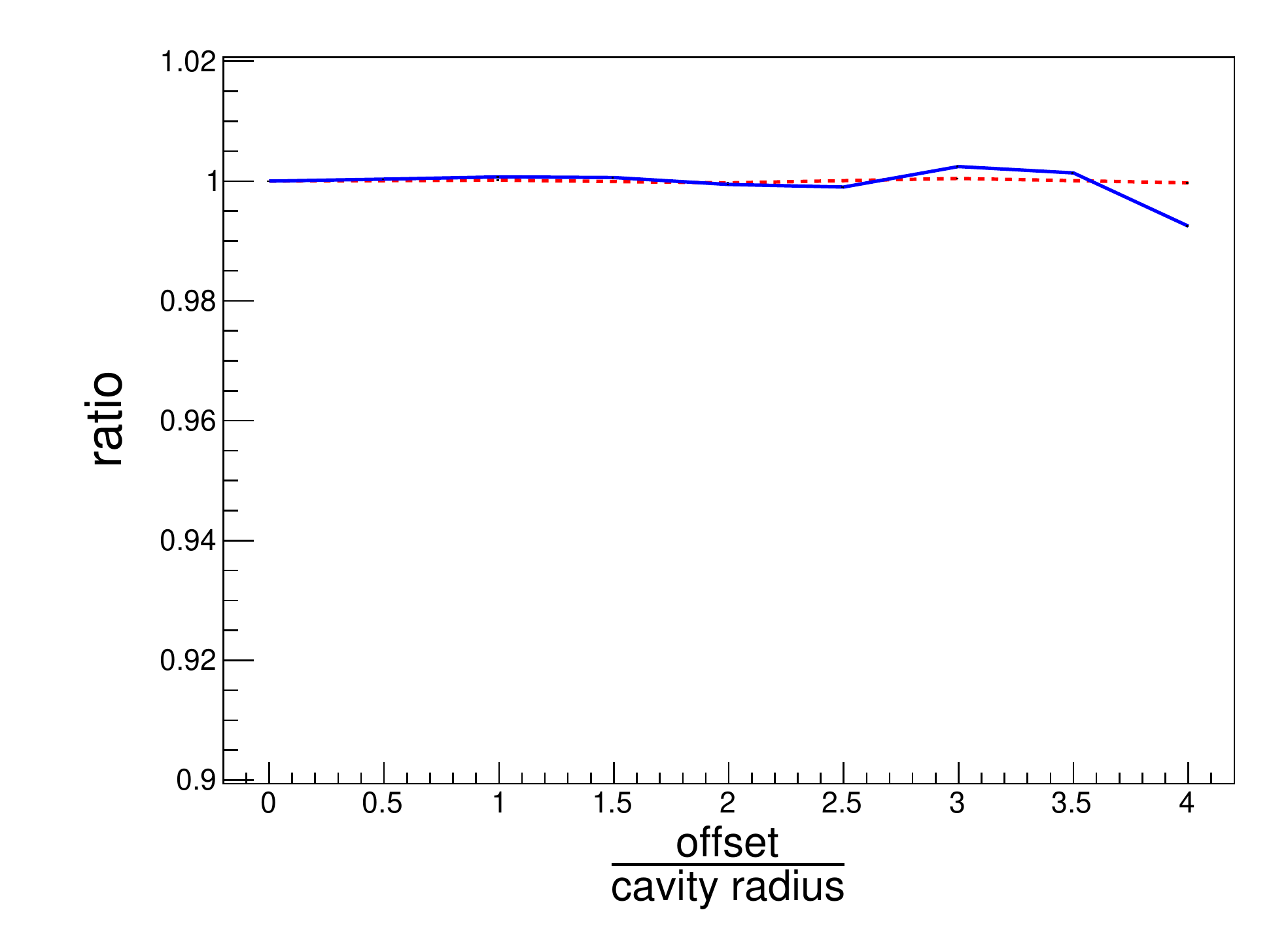}
  \caption{Line (Dots) is ratios of $U_{a,m}$ ($C_B$) to $U^{\rm
      center}_{a,m}$ ($C^{\rm center}_B$) as a function of offset over
    the cavity radius, where $U^{\rm center}_{a,m}$ ($C^{\rm
      center}_B$) is the magnetic energy (magnetic form factor) when
    the cavity is located in the center of the solenoid and offset is
    the distance between the solenoid center and the cavity
    center. The results are calculated with the TM$_{\rm 010}$ mode
    only.}  
  \label{FIG:SIMULATION}
\end{figure}
Figure~\ref{FIG:SIMULATION} shows the numerical demonstration that the
magnetic energy from axion conversions into microwave photons inside a
cylindrical cavity is independent of the position of the cavity inside
the solenoid using the finite element method~\cite{CST}. This
numerical demonstration is calculated with the TM$_{\rm 010}$ mode
only. The three dimensional data points of $\vec{B}_a$ and
$\vec{B}_{ac}$ are simulated for the solenoidal geometry, then the
magnetic energy from the conversion $U_{a,m}$ is calculated from
$\int_V dV\vec{B}_{a}\cdot\vec{B}_{ac}$ as a function of offset over
the cavity radius where the offset is the distance between the
solenoid center and the cavity center. The magnetic form factors given
in Ref.~\cite{EMFF_HOJOO} are also calculated numerically and
superimposed on Fig.~\ref{FIG:SIMULATION}.

\section{Electric and magnetic energy at axion haloscopes with toroidal geometry}
As we pointed out earlier in this paper, the condition
$\vec{\nabla}\times\vec{B}_{\rm external}=0$ is also true for ideal
toroidal magnets. Therefore, $U_{a,e}=U_{a,m}$ regardless of the
position of the toroidal cavity inside the toroidal magnet. This is
very crucial in axion haloscopes with a toroidal magnet because one
cannot define axion signal power without knowing $U_{a,e}$ and
$U_{a,m}$ explicitly. The $U_{a,e}$ of toroidal system fortunately can
be obtained analytically, but one cannot get $U_{a,m}$ of the system
analytically, and thus cannot define axion signal power from axion to
microwave photon conversions with a toroidal magnet. However, since
$U_{a,e}=U_{a,m}$ is also always true for a toroidal system, we do not
have to know $U_{a,m}$ explicitly, instead we can double the $U_{a,e}$
to get the electromagnetic energy from axion to photon conversions
inside the cavity $U_{a,em}$, which is what we experimentally measure
at axion haloscopes.

$U_{a,e}$ and $U_{a,m}$ are proportional to $B^2_{\rm avg}V$,
therefore proportional to the magnetic energy inside the cavity. For a
solenoid, $B^2_{\rm avg}V$ is uniform regardless of the cavity
location, therefore $U_{a,e}$ and $U_{a,m}$ are uniform and
independent of the position of the cavity inside the magnet. For a
toroidal magnet, however, both $B^2_{\rm avg}$ and $V$ vary depending
on the cavity location, therefore $U_{a,e}$ and $U_{a,m}$ depend on
position of the cavity inside the magnet.

\section{Summary}
We have reviewed the electric and magnetic energy in axion
haloscopes. Starting from Maxwell's equations with modifications due
to the axion field, we find that two typical magnet systems satisfying
the condition $\vec{\nabla}\times\vec{B}_{\rm external}=0$ show
$U_{a,e}=U_{a,m}$ regardless of the cavity position inside their
magnets, which refutes the argument reported in
Ref.~\cite{EMFF_HOJOO}. The energy, however, is independent of the
position of the cavity in axion haloscopes with a solenoid, but
depends on those with a toroidal magnet, due to with the $B^2_{\rm
  avg}V$ dependence on it.

\acknowledgments
This work was supported by IBS-R017-D1-2016-a00.


\begin{thebibliography}{99}
\bibitem{HALO_AXION}
  Throughout this paper, our axions are limited to satisfy $\vec{\nabla}a=0$.
\bibitem{COMPLEX}
  Throughout this paper complex electric and magnetic fields are
  implied unless stated otherwise.
\end{thebibliography}
\end{document}